\documentstyle[aps,prl,twocolumn,floats,epsfig]{revtex}
\newcommand{\Z}{{\sf Z \!\!\! Z}}

\newcommand{\Psibar}{\bar{\Psi}}
\newcommand{\p}{\partial}
\begin{document} \draft 
\preprint{DUKE-TH-00-203, MIT-CTP 2961/00}

\title{$SO(10)$ Unification of Color Superconductivity and Chiral Symmetry 
Breaking ?}

\author{Shailesh Chandrasekharan$^\dagger$ and Uwe-Jens Wiese$^\ddagger$}

\address{$^\dagger$ Department of Physics, Box 90305, Duke University,
Durham, NC 27708, U.S.A}
\address{$^\ddagger$ Center for Theoretical Physics,
Massachusetts Institute of Technology,
Cambridge, MA 02139, U.S.A}

\date{March 18, 2000}

\twocolumn[\hsize\textwidth\columnwidth\hsize\csname @twocolumnfalse\endcsname

\maketitle

\begin{abstract}

Motivated by the $SO(5)$ theory of high-temperature superconductivity and 
antiferromagnetism, we ask if an $SO(10)$ theory unifies color 
superconductivity and chiral symmetry breaking in QCD. The transition to the 
color superconducting phase would then be analogous to a spin flop transition. 
While the spin flop transition generically has a unified $SO(3)$ description, 
the $SO(5)$ and $SO(10)$ symmetric fixed points are unstable, at least in 
$(4 - \epsilon)$ dimensions, and require the fine-tuning of one additional
relevant parameter. If QCD is near the $SO(10)$ fixed point, it has interesting
consequences for heavy ion collisions and neutron stars.

\end{abstract} 

]

Grand unified theories (GUT) provide a unified description of the strong and
electroweak interactions at the GUT scale $10^{14}$ GeV \cite{Geo74}. At this 
energy scale all gauge interactions have the same strength, quarks and leptons 
become indistinguishable, and the $SU(3)_c \otimes SU(2)_L \otimes U(1)_Y$ 
gauge symmetry of the standard model is restored to a grand unified group, e.g.
to $SO(10)$. Another type of unification can emerge in critical phenomena. In 
contrast to GUTs where unification arises at short distances, long-range 
collective phenomena may lead to a dynamical enhancement of symmetries at a
critical point. For example, an anisotropic 3-d quantum antiferromagnet with 
$SO(2)_s \otimes \Z(2)$ symmetry has a spin flop transition driven by a 
magnetic field $B$. At small $B$, the staggered magnetization points along the 
easy 3-axis, while at large $B$ it flops into the 12-plane. The first order 
flop transition line ends in a bicritical point from which two second order 
phase transition lines emerge --- one in the 3-d Ising and one in the 3-d XY 
model universality class. At the bicritical point the $SO(2)_s \otimes \Z(2)$ 
symmetry is dynamically enhanced to a unified $SO(3)_s$ symmetry \cite{Fis74}.

Zhang has argued that a similar type of unification may also occur for 
high-temperature superconductors \cite{Zha90}. The undoped precursors of these
materials are quantum antiferromagnets. At low temperature $T$ a staggered 
magnetization is generated which spontaneously breaks the $SO(3)_s$ spin 
rotational symmetry down to $SO(2)_s$. The corresponding Goldstone bosons are 
two antiferromagnetic magnons or spin waves. After doping, i.e. at sufficiently
large chemical potential $\mu$ for the holes, the $SO(3)_s$ symmetry is 
restored and instead the $U(1)_{em}$ gauge group is spontaneously broken by a 
Cooper pair condensate leading to high-temperature superconductivity. When 
treated as a global symmetry, the breaking of $U(1)_{em}$ leads to one massless
Goldstone boson. Once $U(1)_{em}$ is gauged, the Goldstone boson turns into the
longitudinal component of the photon which becomes massive via the 
Anderson-Higgs mechanism. Zhang combined the 3-component staggered 
magnetization vector and the 2-component Cooper pair condensate to an $SO(5)$
``superspin'' vector. In the $SO(5)$ theory, the transition between the 
antiferromagnetic N\'eel phase and the high-temperature superconducting phase 
is a first order ``superspin flop'' transition. At small doping (small $\mu$), 
the superspin lies in the $SO(3)_s/SO(2)_s = S^2 $ easy sphere describing the
staggered magnetization vector. At larger $\mu$, the superspin flops into the 
$U(1)_{em} = S^1$ plane now describing the Cooper pair condensate. The magnons
then turn into massive magnetic modes that persist even in the superconducting 
phase. Indeed, there is experimental evidence for such excitations in 
high-temperature superconductors. The superspin flop transition may end in a 
bicritical point from which two second order lines emerge --- one in the 3-d 
$O(3)$ and one in the 3-d XY model universality class. Zhang has argued that 
the bicritical point has a dynamically enhanced $SO(5)$ symmetry although the
microscopic Hamiltonian is only $U(1)_{em} \otimes SO(3)_s$ invariant.

Here we ask if the $SO(5)$ unified theory of high-temperature superconductivity
and antiferromagnetism can be generalized to an $SO(10)$ unified theory of 
color superconductivity and chiral symmetry breaking in QCD. Although the 
unified group is the same as in a GUT, unification would now occur at 
temperatures around 10 MeV. The analog of the N\'eel phase of a 
high-temperature superconductor precursor at small doping is the chirally 
broken phase of QCD at small baryon chemical potential $\mu$. Here we consider 
QCD with two massless up and down quarks. The chiral symmetry $SU(2)_L \otimes 
SU(2)_R = SO(4)$ then gets spontaneously broken to $SU(2)_{L=R} = SO(3)$ giving
rise to three massless Goldstone pions. As $\mu$ is increased, chiral symmetry 
is restored and one enters the color superconducting phase \cite{Bai84} in 
which a color anti-triplet condensate of quark Cooper pairs leads to the 
spontaneous breaking of $SU(3)_c$ to $SU(2)_c$. As in the $SO(5)$ theory, we
describe the condensate by an effective scalar field. When color is treated as
a global symmetry, its breaking leads to five massless Goldstone bosons in the 
superconducting phase. Once $SU(3)_c$ is gauged, they will get eaten by five 
gluons which become massive via the Anderson-Higgs mechanism. Following Zhang,
we combine the 4-component order parameter for chiral symmetry breaking and the
6-component order parameter for color symmetry breaking to an $SO(10)$ 
``supervector''. In the $SO(10)$ unified theory the transition between the 
chirally broken and the color superconducting phase is a first order 
``supervector flop'' transition. At small $\mu$ the supervector lies in the 
easy 3-sphere $SU(2)_L \otimes SU(2)_R/SU(2)_{L=R} = S^3$ describing the chiral
order parameter. At larger $\mu$ the supervector flops into the 5-sphere 
$SU(3)_c/SU(2)_c = S^5$ now describing the quark Cooper pair condensate. The 
question arises if the supervector flop line can end at a bicritical point 
$(\mu_{bc},T_{bc})$ from which two second order phase transition lines emerge. 
The second order line at $\mu < \mu_{bc}$ separates the chirally broken from 
the high-temperature symmetric phase and is in the universality class of the 
3-d $O(4)$ model. Similarly, the other second order line that separates the 
color superconductor from the symmetric phase is in the universality class of 
the 3-d $O(6)$ model. At a bicritical point the symmetry would be $SO(10)$, not
just $SU(3)_c \otimes SU(2)_L \otimes SU(2)_R \otimes U(1)_B$. However, we will
see that both the $SO(5)$ and the $SO(10)$ symmetric fixed points have one 
additional relevant direction, at least in $(4 - \epsilon)$ dimensions. The 
phase diagram of an $SO(N+M)$ theory is shown in figure 1. This is indeed the 
generic phase diagram for an anisotropic antiferromagnet with $N = 2$ and 
$M = 1$. After fine-tuning the additional relevant parameter, the same phase 
diagram describes high-temperature superconductors ($N = 2$, $M = 3$) and color
superconductors ($N = 6$, $M = 4$) or more precisely high-temperature and color
superfluids because we have not yet gauged $U(1)_{em}$ and $SU(3)_c$.
\begin{figure}[ht]
\begin{center}
\epsfig{file=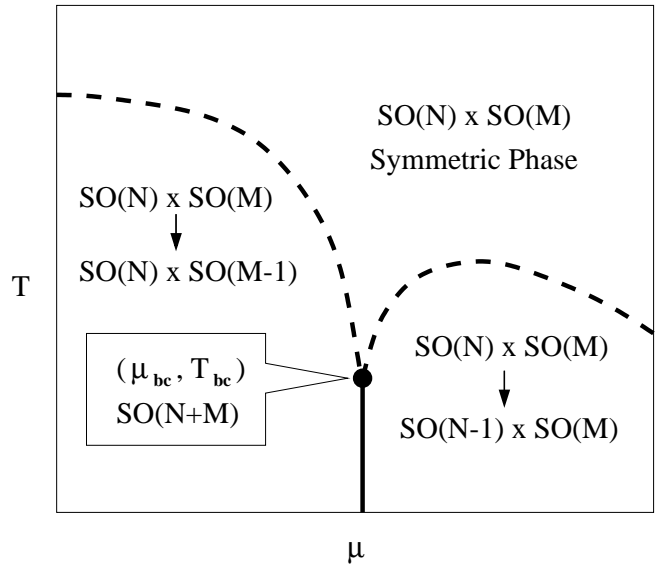,
width=8.6cm,angle=0,
bbllx=0,bblly=0,bburx=390,bbury=334}
\end{center}
\caption{\it Possible phase diagram of an $SO(N) \otimes SO(M)$ theory in the 
$(\mu,T)$ plane. The first order flop transition (solid line) separates two 
phases with symmetry breaking patterns $SO(N) \otimes SO(M) \rightarrow SO(N) 
\otimes SO(M-1)$ and $SO(N) \otimes SO(M) \rightarrow SO(N-1) \otimes SO(M)$. 
It ends in a bicritical point $(\mu_{bc},T_{bc})$ with a dynamically enhanced
$SO(N+M)$ symmetry. Two second order (dashed lines) emerge vertically from this
point.}
\end{figure}

To illustrate the physics of a spin flop transition, let us consider the spin
1/2 anisotropic quantum Heisenberg model on a 3-d cubic lattice in an external 
magnetic field. The corresponding Hamilton operator with nearest neighbor
$\langle x y \rangle$ interactions takes the form
\begin{equation}
H = \sum_{\langle x y \rangle} [J (S_x^1 S_y^1 + S_x^2 S_y^2) + J' S_x^3 S_y^3]
- \vec B \cdot \sum_x \vec S_x.
\end{equation}
We consider antiferromagnetic couplings $J' \geq J > 0$. In the isotropic case
with $J' = J$ and $\vec B = 0$, at low temperatures a staggered magnetization 
vector is dynamically generated, thus spontaneously breaking the $SO(3)_s$ 
symmetry down to $SO(2)_s$. Hence, there are two massless magnons. The 
corresponding low-energy effective theory is formulated in terms of the 
staggered magnetization vector $\vec n = (n^1,n^2,n^3)$ of length 1. In the 
isotropic case the low-energy effective action takes the form
\begin{eqnarray}
S[\vec n]&=&\int_0^{1/T} dt \int d^3x \ \frac{F^2}{2} 
[\p_i \vec n \cdot \p_i \vec n \nonumber \\
&+&\frac{1}{c^2} (\p_0 \vec n + i \vec B \times \vec n) \cdot
(\p_0 \vec n + i \vec B \times \vec n)].
\end{eqnarray}
Here $F^2$ is the spin stiffness and $c$ is the spin wave velocity. The 
magnetic field couples to a non-Abelian conserved charge (the total spin) and 
thus appears as a chemical potential, i.e. as an imaginary non-Abelian constant
vector potential in the Euclidean time direction. To account for an anisotropy 
($J' > J$) we add a potential term $- V_0 (n^3)^2$ to the action that favors 
the 3-direction. With the magnetic field pointing in the 3-direction, the total
potential for constant fields $\vec n$ then takes the form
\begin{equation}
V(\vec n) = - \frac{F^2}{2 c^2} B^2 [(n^1)^2 + (n^2)^2] - V_0 (n^3)^2.
\end{equation}
For small $V_0$ and for $B < B_c = \sqrt{2 V_0 c^2/F^2}$ it is energetically 
favorable for $\vec n$ to point along the easy 3-axis. In this case, only the 
remaining $\Z(2)$ but not the $SO(2)_s$ symmetry is spontaneously broken and 
both magnons pick up a mass. For $B > B_c$, on the other hand, it becomes 
energetically favorable for the staggered magnetization to flop into the 
12-plane. Then the remaining $SO(2)_s$ symmetry gets spontaneously broken 
giving rise to one massless magnon. The spin flop transition is illustrated in 
figure 2.
\begin{figure}[ht]
\begin{center}
\epsfig{file=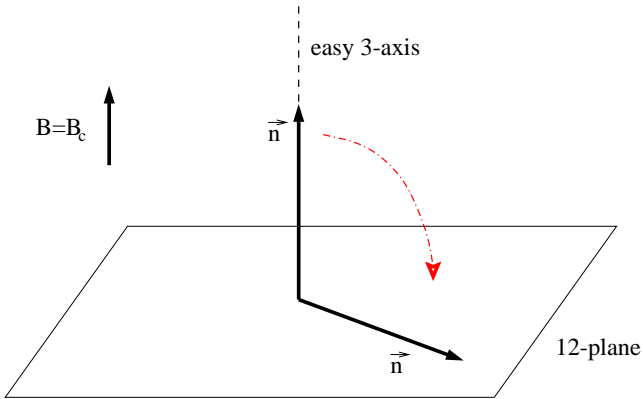,
width=8.6cm,angle=0,
bbllx=0,bblly=0,bburx=476,bbury=291}
\end{center}
\caption{\it The spin flop transition of an anisotropic antiferromagnet in a
magnetic field $\vec B$. For $B < B_c$ the staggered magnetization vector 
$\vec n$ points along the easy 3-axis, and for $B > B_c$ it flops into the 
12-plane.}
\end{figure}
Remarkably, the first order spin flop transition line ends in a bicritical 
point with a dynamically unified $SO(3)_s$ symmetry although the Hamiltonian is
only $SO(2)_s \otimes \Z(2)$ invariant \cite{Fis74}.

Let us now construct the supervector in QCD. We consider left and right-handed 
quark fields $\Psi_L^{f,c}$ and $\Psi_R^{f,c}$ with two flavors $f = 1,2$ and 
three colors $c = 1,2,3$. The chiral symmetry breaking order parameter
\begin{equation}
(\Psibar\Psi)^{fg} = \sum_c \Psibar_L^{f,c} \Psi_R^{g,c}
\end{equation}
is a color singlet, $SU(2)_L$ and $SU(2)_R$ doublet, with baryon number zero. 
The color symmetry breaking order parameter
\begin{equation}
(\Psi\Psi)^c = \sum_{f,g,a,b} \epsilon_{fg} \epsilon_{abc} 
(\Psi_{L,R}^{f,a})^T C \Psi_{L,R}^{g,b},
\end{equation}
on the other hand, is a color anti-triplet, $SU(2)_L \otimes SU(2)_R$ singlet, 
with baryon number $2/3$. Here $T$ denotes the transpose in Dirac space and $C$
is the charge conjugation matrix. The left and right-handed condensates are 
strongly correlated in color space and we represent them by a single scalar
field $(\Psi\Psi)^c$. The two condensates transform differently under $U(1)_A$ 
but we assume that instantons always break that symmetry explicitly. Similarly,
$(\Psibar\Psibar)^c$ is a color triplet, $SU(2)_L \otimes SU(2)_R$ singlet, 
with baryon number $- 2/3$. The group $SO(10)$ contains $SU(3)_c \otimes 
SU(2)_L \otimes SU(2)_R \otimes U(1)_B$ as a subgroup. The 10-dimensional 
vector representation of $SO(10)$ decomposes into
\begin{equation}
\{10\} = \{1,2,2\}_0 \oplus \{\bar 3,1,1\}_{2/3} \oplus \{3,1,1\}_{-2/3},
\end{equation}
and thus naturally hosts the order parameters for chiral symmetry breaking and
color superconductivity. This suggests to construct the 10-component 
supervector $\vec n = (n^1,n^2,...,n^{10})$ with 
\begin{eqnarray}
&&n^c = (\Psi\Psi)^c + (\Psibar\Psibar)^c, \
n^{c+3} = - i [(\Psi\Psi)^c - (\Psibar\Psibar)^c], \nonumber \\
&&n^7 = (\Psibar\Psi)^{11} + (\Psibar\Psi)^{22}, \
n^8 = - i [(\Psibar\Psi)^{12} + (\Psibar\Psi)^{21}], \nonumber \\ 
&&n^9 = (\Psibar\Psi)^{12} - (\Psibar\Psi)^{21}, \
n^{10} = - i [(\Psibar\Psi)^{11} - (\Psibar\Psi)^{22}].\!\!\!
\end{eqnarray}
In the chirally broken phase the 4-component vector $(n^7,n^8,n^9,n^{10})$ 
develops an expectation value, thus breaking $SU(2)_L \otimes SU(2)_R$ 
spontaneously to $SU(2)_{L=R}$. The corresponding Goldstone pions are described
by fields in the $SU(2)_L \otimes SU(2)_R/SU(2)_{L=R} = S^3$ easy 3-sphere. In 
the color superconducting phase, on the other hand, the 6-component vector 
$(n^1,n^2,...,n^6)$ gets an expectation value and the supervector flops into 
the 5-sphere $SU(3)_c/SU(2)_c = S^5$ that parameterizes the corresponding five 
massless Goldstone bosons.

In analogy to the antiferromagnet discussed before, we now consider a unified
theory with symmetry breaking pattern $SO(N+M) \rightarrow SO(N+M-1)$. The
corresponding Goldstone bosons are described by an $(N + M)$-component unit 
vector $\vec n$. In the absence of $SO(N+M)$ symmetry breaking terms (other 
than the chemical potential), the low-energy effective action takes the form
\begin{eqnarray}
S[\vec n]&=&\int_0^{1/T} dt \int d^3x \ \frac{F^2}{2} 
[\p_i n^\alpha \p_i n^\alpha \nonumber \\
&+&\frac{1}{c^2} (\p_0 n^\alpha + A_0^{\alpha\beta} n^\beta)
(\p_0 n^\alpha + A_0^{\alpha\gamma} n^\gamma)].
\end{eqnarray}
As before, the chemical potential $\mu$ couples as an imaginary non-Abelian 
constant vector potential
\begin{equation}
A_0^{\alpha\beta} = i \mu \sum_{c = 1,...,N/2} 
(\delta^{\alpha,c} \delta^{c+N/2,\beta} - 
\delta^{\alpha,c+N/2} \delta^{c,\beta})
\end{equation} 
in the Euclidean time direction. As for the anisotropic antiferromagnet with 
$N = 2$ and $M = 1$, we introduce explicit symmetry breaking terms that reduce 
the symmetry to $SO(N) \otimes SO(M)$. The cases $N = 2, M = 3$ and $N = 6,
M = 4$ correspond to high-temperature and color superconductors, respectively.
We add a potential term $- V_0 [(n^{N+1})^2 + ... + (n^{N+M})^2]$ to the action
that favors the easy $(M-1)$-sphere. For QCD this leads to chiral symmetry 
breaking. Actually, the above term breaks the $SO(10)$ symmetry down only to 
$SO(6) \otimes SO(4) = SU(4) \otimes SU(2)_L \otimes SU(2)_R$, not to $SU(3)_c 
\otimes SU(2)_L \otimes SU(2)_R \otimes U(1)_B$. This is sufficient because at 
low energies one cannot distinguish between the symmetry breaking patterns 
$SU(3)_c \otimes U(1)_B \rightarrow SU(2)_c \otimes U(1)_B$ and $SO(6) 
\rightarrow SO(5)$ in $(4 - \epsilon)$ and even in $(2 + \epsilon)$ dimensions 
\cite{Fri85}. The total potential for constant fields $\vec n$ then takes the 
form
\begin{eqnarray}
V(\vec n)&=&- \frac{F^2}{2 c^2} \mu^2 [(n^1)^2 + ... + (n^N)^2] \nonumber \\
&-&V_0 [(n^{N+1})^2 + ... + (n^{N+M})^2].
\end{eqnarray}
For $\mu < \mu_c = \sqrt{2 V_0 c^2/F^2}$ it is energetically favorable for the 
supervector $\vec n$ to lie in the easy $(M-1)$-sphere. For $\mu > \mu_c$, on 
the other hand, the supervector flops into the $(N-1)$-sphere. More generally,
one should take into account symmetry breaking effects in $F$ and $c$ as well. 

To investigate if an $SO(N+M)$ unified theory can describe the phase 
transition in a high-temperature or color superconductor, we investigate the 
renormalization group flow equations for $SO(N) \otimes SO(M)$ invariant scalar
field theories in $(4 - \epsilon)$ dimensions with a potential
\begin{equation}
V(\vec \Phi) = \frac{1}{4!}[g_1 (\Phi_N^2)^2 + g_2 (\Phi_M^2)^2 + 
2 g_3 \Phi_N^2 \Phi_M^2].
\end{equation}
Here $\Phi_N^2 = (\Phi^1)^2 + ... + (\Phi^N)^2$ and $\Phi_M^2 = (\Phi^{N+1})^2 
+ ... + (\Phi^{N+M})^2$. In agreement with \cite{Kos76}, we obtain the 
$\beta$-functions for the couplings $g_1$, $g_2$ and $g_3$ as
\begin{eqnarray}
\beta_1&=&- \epsilon g_1 + [(N+8) g_1^2 + M g_3^2]/6, 
\nonumber \\
\beta_2&=&- \epsilon g_2 + [(M+8) g_2^2 + N g_3^2]/6, 
\nonumber \\
\beta_3&=&- \epsilon g_3 + g_3 [(N+2) g_1 + (M+2) g_2 + 4 g_3]/6.
\end{eqnarray}
There are three distinct fixed points. The $SO(N+M)$ invariant fixed point
$g_1 = g_2 = g_3 = 6 \epsilon/(N+M+8)$ is stable only for $N+M < 4$. This is
the case for the anisotropic antiferromagnet, but not for high-temperature or
color superconductors. For high-temperature superconductors ($N = 2$, $M = 3$) 
a biconical fixed point with $g_1 = 0.5429 \epsilon$, $g_2 = 0.5085 \epsilon$, 
$g_3 = 0.3215 \epsilon$ is stable. Finally, a decoupled fixed point $g_1 = 6 
\epsilon/(N+8)$, $g_2 = 6 \epsilon/(M+8)$, $g_3 = 0$ is stable for $NM + 2(N+M)
> 32$, which is the case for QCD ($N = 6$, $M = 4$). Both for high-temperature
and for color superconductors the $SO(N+M)$ invariant fixed point is unstable
against perturbations in an additional relevant direction. This suggests that, 
unlike anisotropic antiferromagnets, high-temperature and color superconductors
generically do not enhance their symmetries to $SO(N+M)$ at a bicritical point.
Still, after fine-tuning the additional relevant parameter (e.g. the strange 
quark mass in QCD), one can reach the $SO(N+M)$ symmetric point. Without 
fine-tuning, the phase diagram may look as suggested in \cite{Pis99}. However,
it is not clear if the calculation in $(4 - \epsilon)$ dimensions correctly 
describes the physics in three dimensions. A detailed numerical study of a 3-d 
$SO(6) \otimes SO(4)$ model could clarify this question.

If QCD is close enough to the $SO(10)$ symmetric point, it is interesting to 
ask if the supervector can play a dynamical role in nature. First of all, when 
the strange quark is introduced, a new color superconducting phase with 
color-flavor locking arises \cite{Alf99}. This phase may be analytically 
connected to the ordinary hadronic phase \cite{Sch99}. In that case, there can 
be no supervector flop transition. However, when the strange quark is 
sufficiently heavy, a flop transition may exist. When the mass of the up and 
down quarks is taken into account, the second order line at $\mu < \mu_{bc}$ 
turns into a crossover and the point $(\mu_{bc},T_{bc})$ becomes a tricritical 
point. Furthermore, when $SU(3)_c$ is gauged, the Goldstone bosons in the color
superconducting phase get eaten and the other second order line at 
$\mu > \mu_{bc}$ may also turn into a crossover. In that case we are left with 
a single first order line with a critical end point at $(\mu_{bc},T_{bc})$. 
That point is in the universality class of the 3-d Ising model with the sigma 
mode as the only remaining massless excitation. It has been argued that the 
critical end point may be detectable through event by event fluctuations in 
heavy ion collisions \cite{Ste98}. In the $SO(10)$ theory this point is tied to
the color superconducting phase and would thus be in a region that is very hard
to probe with heavy ion collisions. If this property of the $SO(10)$ theory 
persists for QCD, heavy ion collisions can reach the quark-gluon plasma only 
through a smooth crossover. On the other hand, if $T_{bc}$ is very small, 
neutron star cores may be close to an $SO(10)$ invariant quantum critical point
with unusually light modes analogous to the ones observed in high-temperature 
superconductors. In any case, both in color and in high-temperature 
superconductors it is natural to think about unification far below the GUT 
scale.

\section*{Acknowledgements}

We are indebted to D. H. Friedan for extremely helpful advice and to D. Kaplan 
and K. Rajagopal for clarifying and stimulating discussions. We also like to 
thank M. Alford and S.-C. Zhang for interesting discussions about color and 
high-temperature superconductivity. We gratefully acknowledge the hospitality 
of the INT in Seattle where this work was done. Our work is supported in part 
by funds provided by the U.S. Department of Energy (D.O.E.) under cooperative 
research agreements DE-FC02-94ER40818 and DE-FG02-96ER40945. U.-J. W. also 
acknowledges the support of the A. P. Sloan foundation.

\end{document}